\documentclass[a4paper]{jpconf}
\usepackage{graphicx}
\usepackage{color}
\usepackage{amsmath}

\def\jpb{{\em J. Phys. B: At. Mol. Opt. Phys.}~}
\def\pra{{\em Phys. Rev. A}~}

\def\prl{{\em Phys. Rev. Lett.}~}

\def\jetp{{\em J. Exp. Theor. Phys.}~}
\def\etal{{\em et al. }}

\newcommand{\vecr}{\mathbf{r}}

\newcommand{\vecA}{\mathbf{A}}

\newcommand{\vecE}{\mathbf{E}}

\newcommand{\beq}{\begin{equation}}
\newcommand{\eeq}{\end{equation}}

\def\cmcmcm{\mathrm{cm^{-3}}}
\def\Wcmcm{\mathrm{W/cm^2}}
\def\ds{\displaystyle}
\def\mum{\mathrm{\mu m}}
\def\fs{\mathrm{fs}}

\begin{document}
\title{Bright single-cycle terahertz source based on gas cells irradiated by two-color laser pulses}

\author{T.V. Liseykina}

\address{Institute of Physics, University of Rostock, D-18051 Rostock, Germany}
\address{Institute of Computational Mathematics and Mathematical Geophysics SB RAS, Lavrent'ev~ave. 6, 630090 Novosibirsk, Russia}

\author{S.V. Popruzhenko}

\address{Prokhorov General Physics Institute RAS, Vavilova str. 38, 119991 Moscow, Russia}
\address{Voronezh State University, Universitetskaya pl. 1, 394036 Voronezh, Russia}

\ead{sergey.popruzhenko@gmail.com}

\begin{abstract}
We study the excitation of electron currents in a transparent cell of sub-millimeter size filled by an atomic gas and illuminated by an intense two-color femtosecond laser pulse.
The pulse consists of a strong fundamental component and its second harmonic of low intensity, both circularly polarized. 
We show that for a sufficiently small $20~\mum$-size interaction volume the plasma oscillation excited by asymmetric ionization is almost spatially homogeneous.
This coherent dipole plasma oscillation results in a remarkably efficient conversion of the electron energy into that of radiation emitted in the terahertz frequency domain.
Simultaneously, strong quasi-static electric fields of maximal strength $E_\mathrm{m}\simeq 10~\mathrm{MV/cm}$ are shown to exist inside the plasma during several hundred femtoseconds after the ionizing two-color laser pulse has gone.
\end{abstract}

\section{Introduction}
Electromagnetic radiation in the terahertz (THz) frequency domain is known to have a considerable potential for utilization both in fundamental experimental physics and in various applications.
This includes THz spectroscopy of molecular gases and solids, control of magnetism in complex materials, noninvasive diagnostics, imaging and more, see \cite{biomed-rev-jpd06,lee-book,dexheimer-book,THz-rev-opt16,THz-rev-jpd17} for review and references.
Sources of high-intensity short THz pulses are of particular interest as they provide nonlinear regime of interaction and can be used for control of fast processes.
Several efficient methods for the generation of strong THz waves are presently known and being applied in laboratories \cite{THz-rev-opt16,THz-rev-jpd17}.
Optical rectification in crystals \cite{yeh-apl07} provides the currently highest convergence efficiency from optical or infrared (IR) to THz radiation at frequencies $\nu<3~\mathrm{THz}$.
Such sources can deliver THz pulses of $\simeq 10~\mu\mathrm{J}$ energy and $\simeq 1~\mathrm{MV/cm}$ electric field strength. 
Their main limitations are the relatively narrow spectral width determined by material absorption and the intensity damage threshold of crystals which becomes particularly restrictive at high repetition rates of pump lasers.
Another family of perspective methods for the THz generation is based on laser frequency conversion in plasmas.
This includes in particular emission of relativistic laser plasmas at solid-state densities \cite{gopal-njp12,gopal-prl13} and several schemes based on nonlinear ionization of gases \cite{bartel-ol05,kim-oe07,vved-prl09,berge-prl13,vved-prl18}.
These ionization schemes are technically simple and can employ ambient air at normal conditions as a target.
The emitted spectrum is broad and typically extends up to $10-30~\mathrm{THz}$.
In combination with the absence of a damage threshold and the possibility of using high-repetition pump lasers, this makes ionization-based sources of THz radiation a promising alternative to nonlinear crystals.

Two-color laser radiation consisting of a strong pulse of frequency $\omega$ and its relatively weak second harmonic (SH) is conventionally used to excite asymmetric ionization currents emitting THz waves  \cite{kim-oe07,you-prl12}. 
By now, employing linearly polarized 800-nm pulses and their co-polarized SH, THz waves with an electric field strength of the order of 10 MV/cm \cite{kim-apl14} have been obtained providing the presently strongest THz source operating in the high-repetition-rate regime. 
Over the past years the $\omega-2\omega$ scheme has been extensively studied employing different field polarization states and wavelengths \cite{dai-prl09,wen-prl09,clerici-prl13,meng-apl16}.
According to recent theoretical predictions \cite{fedorov-pra18,tulsky-pra18}, application of mid-infrared two-color pulses instead of conventionally used 800-nm radiation may help increasing the THz emitted energy by one order of magnitude or even more.
This foreseeing enhancement can make THz sources of interest for fundamental experiments in atomic and molecular physics and for other applications requiring nonlinear regimes of interaction. 

The physical mechanism of $\omega-2\omega$ THz generation is well understood on the single-atom level. A two-color laser field generally generates asymmetric photoelectron momentum distributions, so that after the averaging over fast oscillations with the frequencies $\omega$ and $2\omega$ and over the distribution, the electron current does not vanish, in contrast to the case of a quasi-monochromatic field \cite{kim-pp09,kotelnikov-jetp11,kim-pra13,poprz-pra15}.
The collective response of media ionized by bichromatic fields is less studied owing to a high complexity of the problem.
Most of the theory developed so far used simple models for the photo-induced current or nonlinear susceptibility which enter the right-hand side of an exact or reduced inhomogeneous wave equation (see examples in \cite{vved-prl18,kim-pp09,kim-pra13}).
Such approaches allow to compute the response of spatially extended media including filaments, but they suffer from model simplifications and do not take self-consistently into account the back reaction of THz radiation created in the plasma on the electron motion.
Meanwhile, radiation can significantly influence the electron dynamics as the number of coherently emitting electrons may reach giant values of $\ds\sim 10^{10}$ and higher.

In this paper, we address the problem of the macroscopic THz response of laser-driven plasma in a complimentary way by employing a particle-in-cell (PIC) simulation of the plasma dynamics during and after the interaction with a two-color ionizing laser pulse.
Our simulation includes the ionization step and allows for a fully self-consistent calculation of the electron current including the back reaction of the coherently emitted radiation on the plasma dynamics.
We employ the PIC code UMKA developed for studying laser-plasma interactions in the strong-field regime and adapted to include ionization \cite{Vshivkov-1998,droplet-2013}.
High numerical costs of PIC simulations restrict the interaction volume in all dimensions by size $\sim 100~\mum$, so that instead of considering filaments created in open air we examine radiation emitted from a 
spatially restricted target.
Such targets can be realized with thin gas jets or small gas-filled cells, the latter were used in experiment \cite{meng-apl16}.
We analyze the plasma dynamics, calculate distributions of the quasi-static electric field inside the plasma and the radiation spectra emitted in the forward direction, and estimate the efficiency of the electron energy conversion into that of THz radiation.
Our results show a relatively high energy conversion which appears more efficient for a smaller interaction volume and indicate the presence of strong long-living quasi-static electric fields.
On this basis, we suggest that the application of small gas-filled cells or thin gas beams where a two-color laser field can induce an almost homogeneous oscillating dipole, may lead to a much higher IR-to-THz energy conversion efficiency than that presently achieved in the regime of filamentation.

\section{Basic equations and numerical model}
We consider a two-color circularly polarized pulse described by the vector potential  
\beq
\vecA(\vecr,t)=\vecA_{\omega}(\vecr,t)+\vecA_{2\omega}(\vecr,t)
\label{A}
\eeq
whose form is determined by its time dependence in the plane $(x=0,y,z)$:
\beq
\vecA_{\omega}(x=0,y,t)=\frac{E_0}{\omega}\e^{-\frac{y^2}{2R^2}}\bigg(0,f(\varphi)\cos(\varphi),f(\varphi-\pi)\sin(\varphi)\bigg)~,~~f(\varphi)=\sin^2\bigg(\frac{\varphi}{2N}\bigg)~.
\label{A1}
\eeq
Here $E_0$ and $\omega$ are the electric field amplitude and the carrier frequency of the fundamental pulse with duration of $N$ periods, $R=16~\mum$ corresponding to the focal spot diameter (FWHM) of $27~\mum$ and the phase $\varphi=\omega(t-x/c)$ with c being the speed of light.
The phase variable $\varphi\in [0,2\pi N]$, and $\vecA=0$ outside of this interval.
The SH field has the same spatio-temporal envelope
\beq
\vecA_{2\omega}(x=0,y,t)=\epsilon\frac{E_0}{2\omega}\e^{-\frac{y^2}{2R^2}}\bigg(0,f(\varphi)\cos(2\varphi-\alpha),f(\varphi-\pi)\sin(2\varphi-\alpha)\bigg)~
\label{A2}
\eeq
and is shifted by phase $\alpha$ with respect to the fundamental.
The fundamental field amplitude was taken $E_0=0.17~\mathrm{at.u.}=0.87~\mathrm{GV/cm}$, corresponding to intensity 
$\ds\mathcal{I}=2\times 10^{15}~\Wcmcm$ of circularly polarized radiation.
The number of cycles $N=100$ for $0.8~\mum$ wavelength and $N=40$ for $2~\mum$ corresponds to the fixed pulse duration $\tau=267~\fs$.
The relative electric field amplitude in Eq.~\eqref{A2} is taken $\epsilon=0.22$ corresponding to a 5\% relative intensity of SH.
The choice of circularly polarized radiation is based on its higher efficiency in the excitation of the net photoelectron current responsible for emission of THz waves \cite{meng-apl16,tulsky-pra18}.
For an isotropic gas target, the phase shift $\alpha$ between the two fields in Eqs.~\eqref{A1} and \eqref{A2}, determines the orientation of the symmetry axis in the polarization plane, but otherwise does not affect the plasma dynamics \cite{tulsky-pra18}.
In our calculations 
we use a rectangular two-dimensional cell which makes results $\alpha$-dependent.
To focus on the physically relevant situation when plasma oscillations are exited along the direction where the plasma size is spatially restricted, we adjust the value of $\alpha$ to direct the net photoelectron momentum after ionization along the $y$ axis.
In the case of linearly polarized pulses, the plasma dynamics is expected to be generally similar, but with a clear dependence on the phase shift, as was observed in \cite{meng-apl16}.

The rectangular gas cell is restricted in the $(x,y)$ plane, $0\le x\le L, -L/2\le y\le +L/2,$ with $L=20~\mum$ or $50~\mum$ and homogeneous in $z$ direction which 
makes the calculation two-dimensional (2D) in the position space, while electron momenta and electromagnetic fields are calculated in the full dimension.
The incoming laser pulse propagating along the $x$ axis is determined by Eqs.~\eqref{A1}, \eqref{A2} on the front side of the cell $x=0,$
while the electromagnetic field in the whole space is calculated numerically from the system of Vlasov-Maxwell equations. 
With the phase shift $\alpha$ fixed as described above, translation invariance along the $z$ axis is not expected to introduce unphysical effects.
Spatial and temporal resolution were $\ds\Delta x =\Delta y=(2\pi c/\omega)/40$  and $\ds\Delta t=(2\pi/\omega)/80,$ correspondingly, and  $64$ macroparticles per species per numerical cell were used. 

The gas is assumed initially neutral, and its ionization proceeds along the field tunneling mechanism.
We describe ionization events probabilistically, with the distribution function determined by the tunneling ionization rate taken from \cite{popov-usp04,poprz-jpb14}.
We only consider single-electron ionization, so that all the atomic species are either neutral or single-charged.
When an ionization event happens, the charge state of a chosen ion changes from $0$ to $+1$, and a free electron is created at rest at the position of the ion. 
The energy required for ionization is subtracted from the field via the work of the "ionization current" $\mathbf{j}_{\mathrm{ion}}$ parallel to the electric field at the ion location. 
The energy conservation is secured by the condition that the value $(\mathbf{j}_{\mathrm{ion}}\cdot\mathbf{E})\Delta t$ is equal to the energy spent on ionization per time step $\Delta t$ \cite{Rae,Mulser-Cornolti-Bauer}. 
If the field energy in a cell is insufficient for further ionization, this cell is not considered anymore during the current time step \cite{Kemp-2004}. 
For target and laser parameters we consider below, a small fraction of the laser energy is sufficient to singly ionize all atoms in the interaction volume, so that effects of saturation play a minor role.
In order to check the role of cell boundaries we have also made several runs for bigger cells with the sizes up to $150~\mum$ in the longitudinal and lateral directions.
Results of these calculations show that the effective lateral size of the plasma is determined by that of the focal waist $\simeq 30~\mum$ and does not depend on the cell width and on its shape in the lateral direction.
Instead, the longitudinal cell size has a direct effect on the plasma response.

\section{Numerical results}
\begin{figure}
\begin{center}
\leftline{\includegraphics[height=10cm]{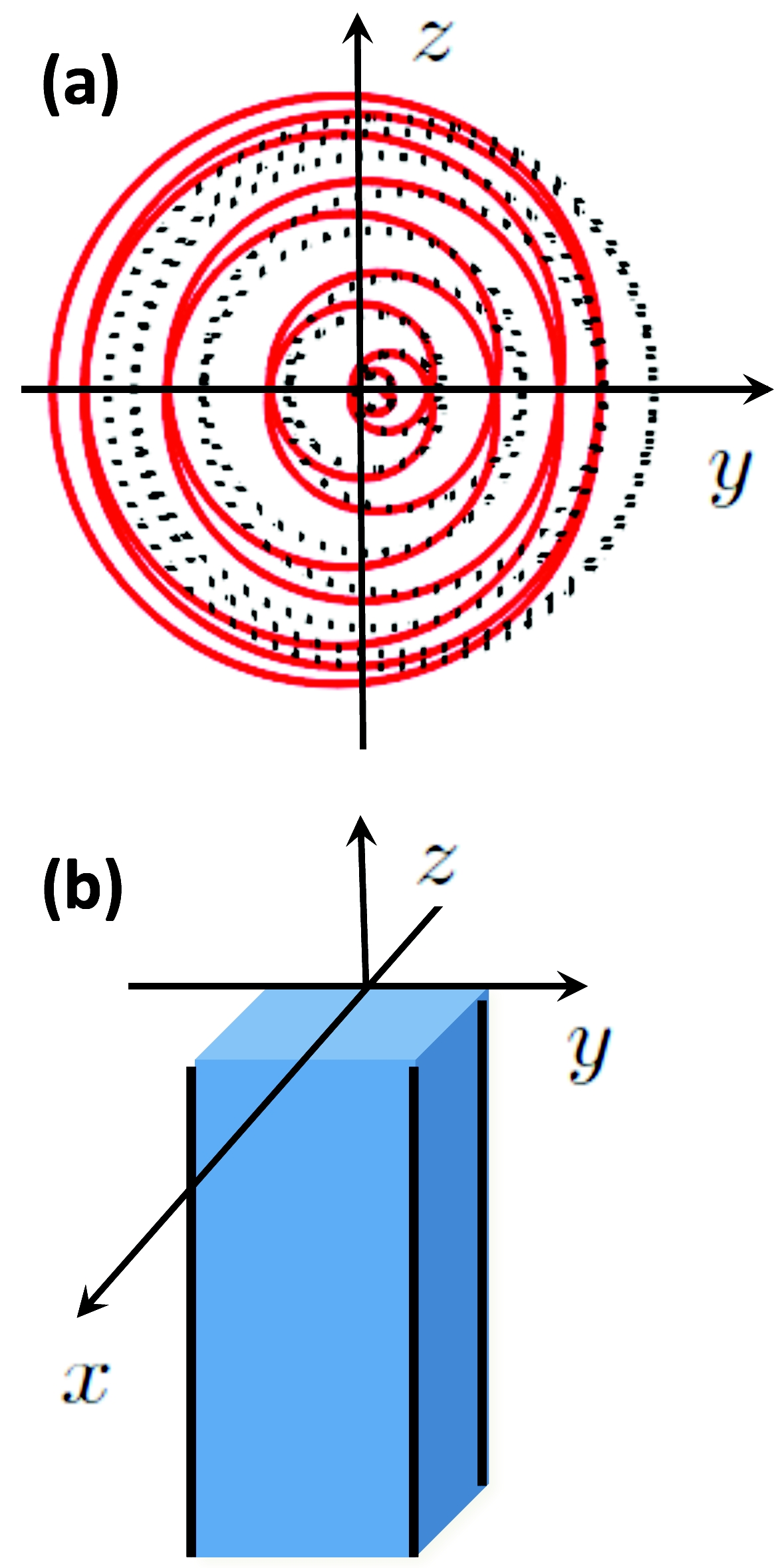}
\includegraphics[height=10cm]{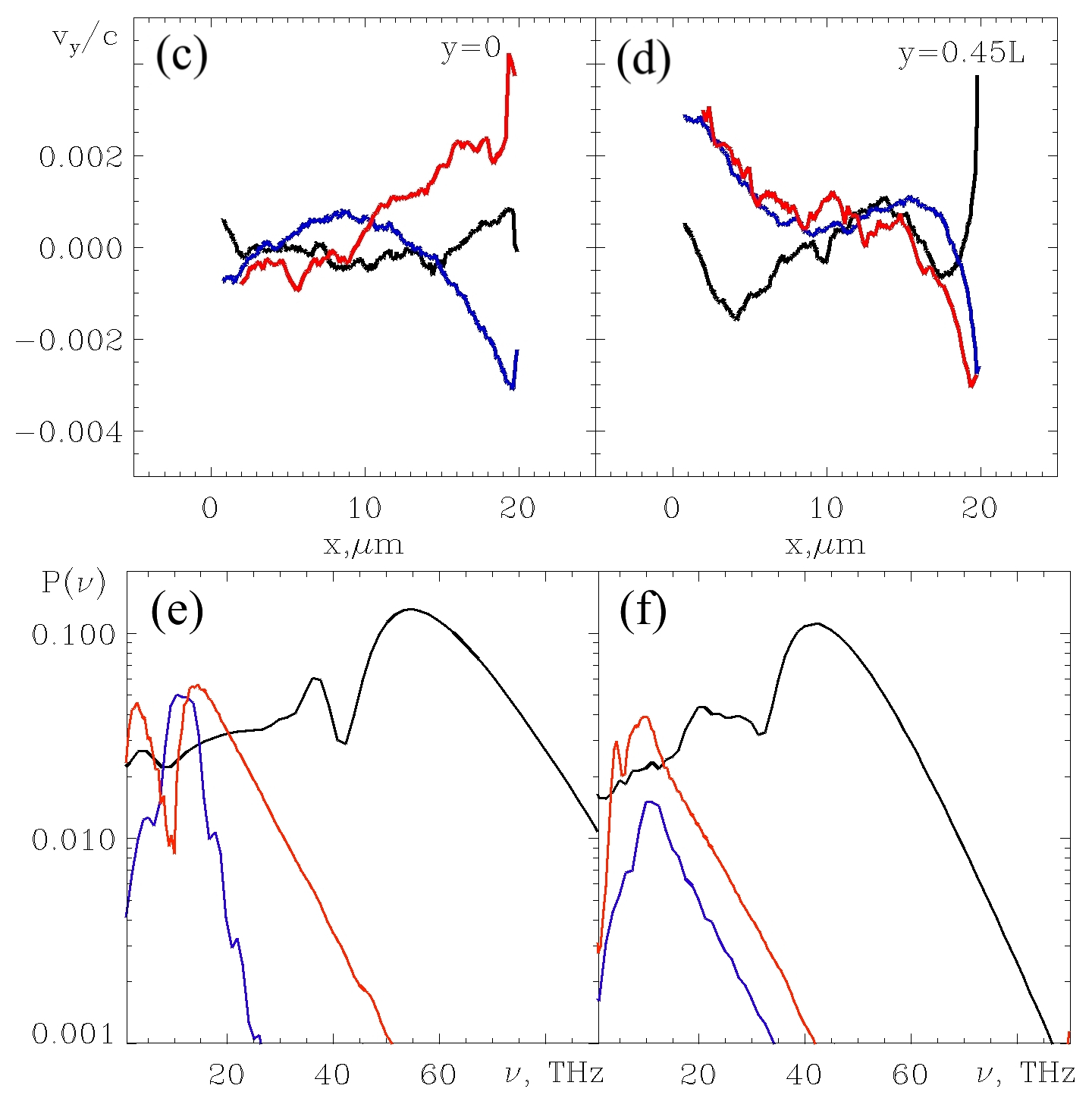}}
\end{center}
\caption{(color online) Vector potentials of the single-color $\vecA_{\omega}(0,0,t)$ (black dotted line) and the two-color $\vecA(0,0,t)$ (solid red line) laser pulses defined by Eqs.~\eqref{A}, \eqref{A1} and \eqref{A2} (a); geometry of the 2D target (b); distributions in the electron velocity $v_y(x)$ normalized to the speed of light $c$ at $t=333~\fs$ and $y=0$ (c) and $y=0.45L$ (d) for the $20~\mum$ cell; spectral radiation power $P(\nu)$ in the forward direction measured in relative units for the $50~\mum$ (e) and $20~\mum$ (f) cells. Black lines correspond to parameters $(\lambda=0.8~\mum),~n_0=10^{19}~\cmcmcm$, blue lines -- to $(\lambda=0.8~\mum),~n_0=10^{18}~\cmcmcm$ and red lines -- to $(\lambda=2~\mum),~n_0=10^{18}~\cmcmcm$.
On panel (a), the number of cycles in the pulse is reduced to $N=10$ for better visibility.}
\label{fig:fig1}
\end{figure}

Figure \ref{fig:fig1} shows the vector potential, Eq.~\eqref{A}, the target geometry, distributions in electron velocity $v_y$ taken at a time instant $t=333~\fs$ approximately equal to the time when the laser pulse leaves the cell after the interaction, and spectra $P(\nu)\sim\vecE(\nu)\vecE^*(\nu)$ of radiation emitted in the forward direction.
These distributions demonstrate a clear dependence on the cell size, on the laser wavelength, and on the gas density.
Firstly, at intensity $\mathcal{I}=2\times 10^{15}~\Wcmcm$, single-electron tunnel ionization of argon happens within a couple of optical cycles on the front edge of the pulse, so that during the interaction and after the laser pulse is gone the electron concentration of the plasma $n_e$ is close to that of atoms $n_0$. 
As a consequence, the plasma frequency $\ds\omega_p=\sqrt{4\pi e^2n_e/m_e}$ approximately equals to 
$5\times 10^{13}~\mathrm{s^{-1}}$ for $n_0=10^{18}~\cmcmcm$ and to $1.5\times 10^{14}~\mathrm{s^{-1}}$ for $n_0=10^{19}~\cmcmcm$. 
These numbers agree with the positions of maxima in the radiation spectra.
Note that the 
plasma wavelength, $\ds\lambda_p=2\pi c/\omega_p$, which is equal to $40~\mum$ and $13~\mum$ for $10^{18}~\cmcmcm$ and
$10^{19}~\cmcmcm$, respectively, determines the number of oscillations per cell size $L$ in the electron distribution. 
Secondly, the laser wavelength determines the initial photoelectron velocity (see e.g. \cite{tulsky-pra18} for details) $\ds v_0\approx keE_0\lambda/(2\pi m_ec)$ with $k\approx 0.1$.
These parametric scalings are clearly seen on the distributions shown in Fig.~\ref{fig:fig1}(c-f).

\begin{figure}
\begin{center}
\centerline{\includegraphics[width=0.33\textwidth]{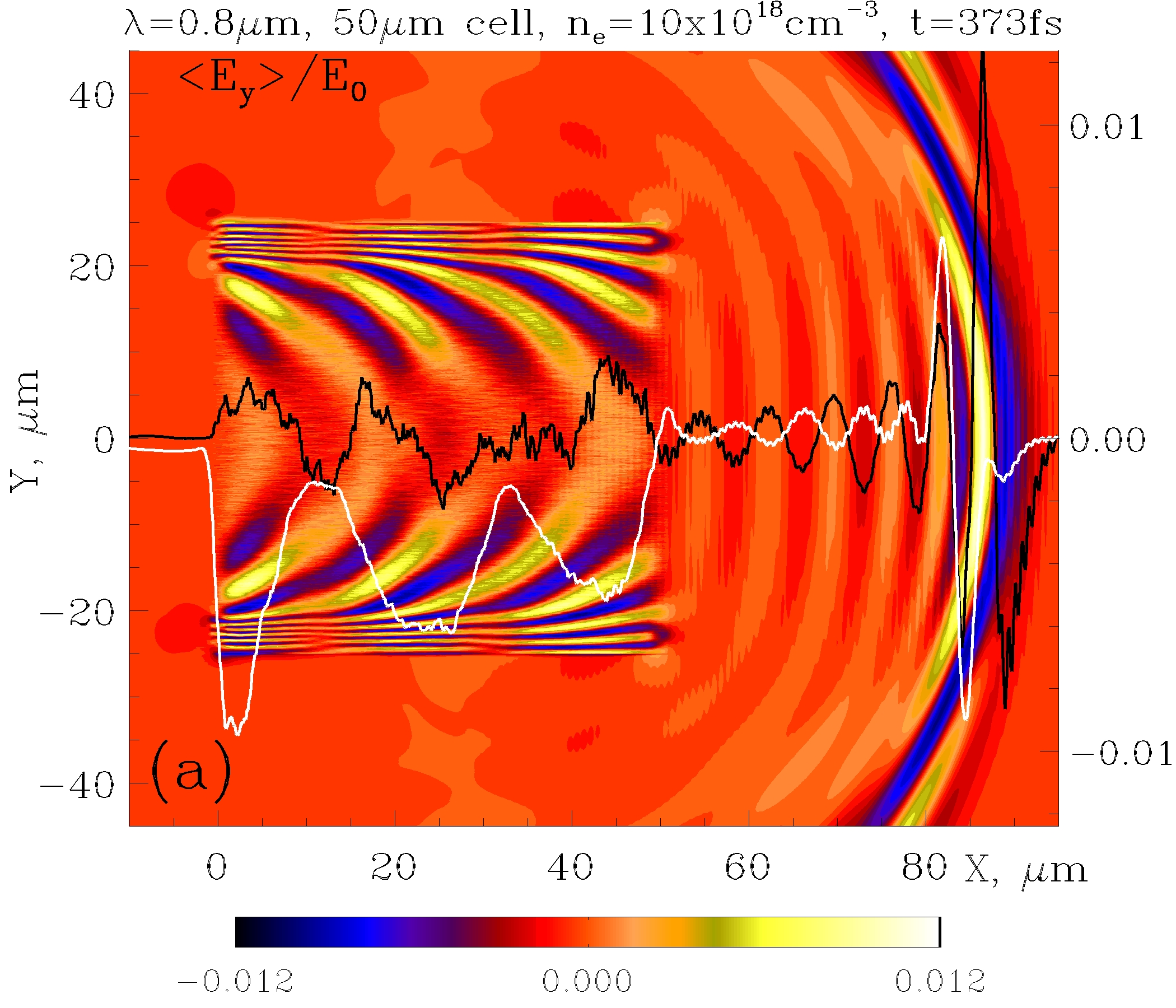}
\includegraphics[width=0.33\textwidth]{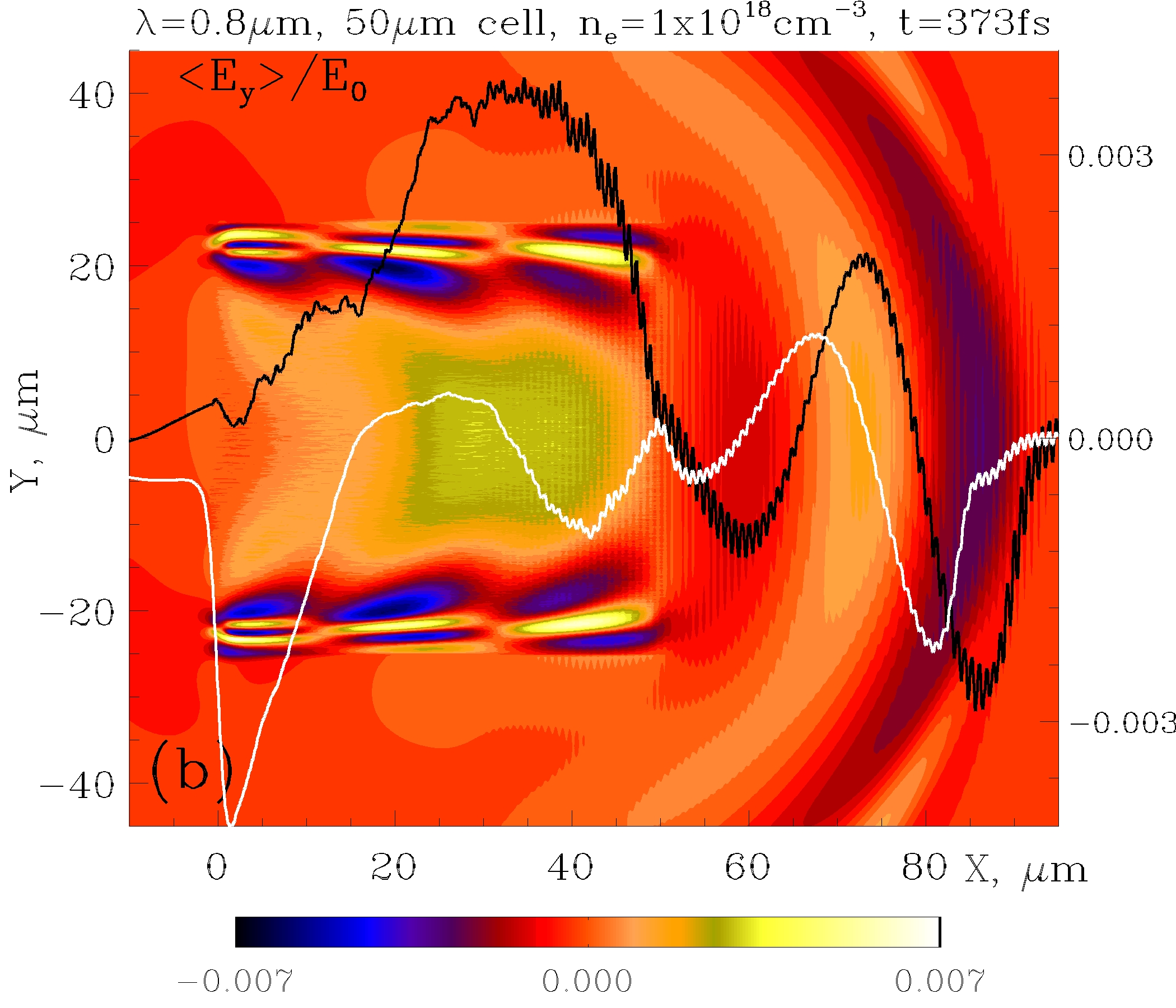}
\includegraphics[width=0.33\textwidth]{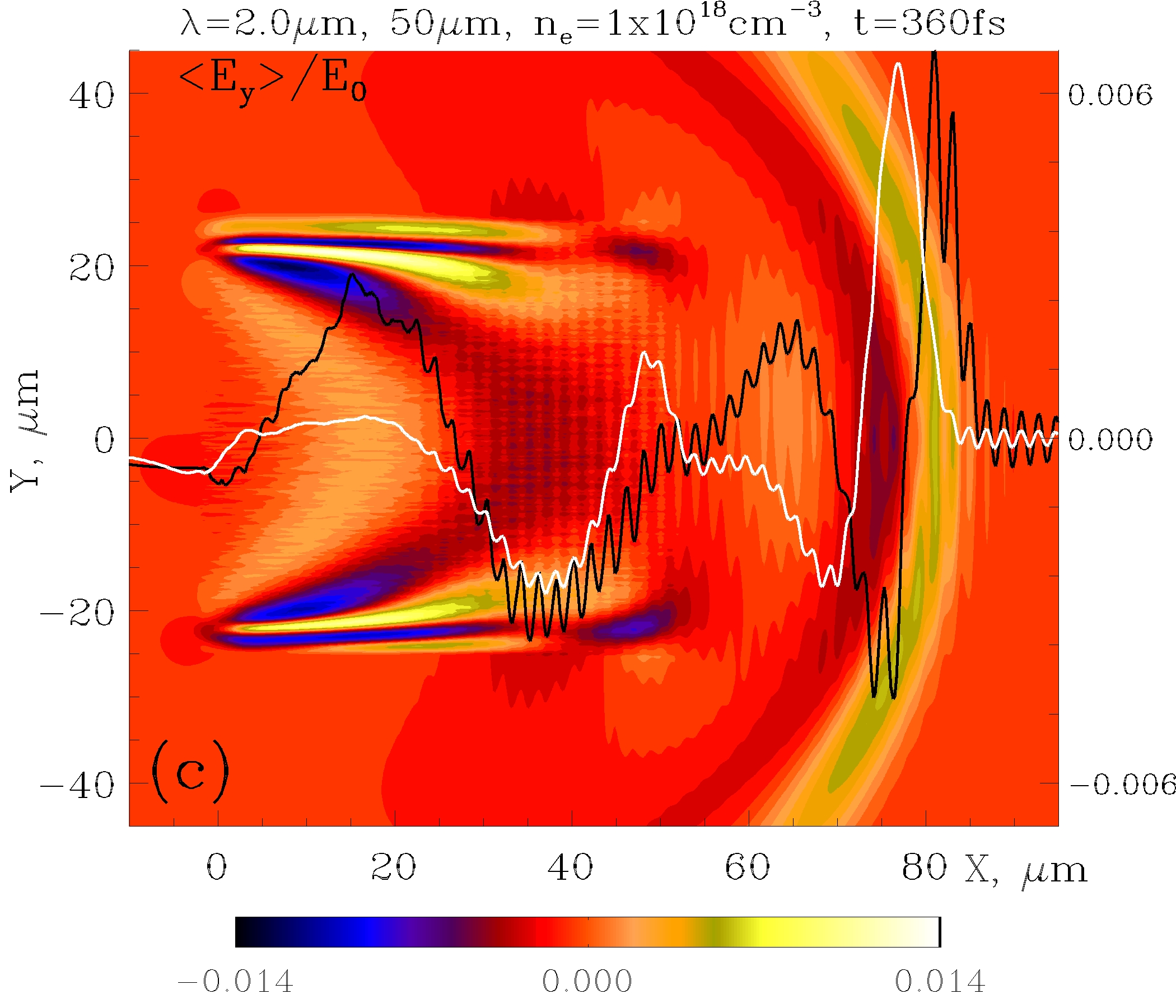}
}
\centerline{\includegraphics[width=0.33\textwidth]{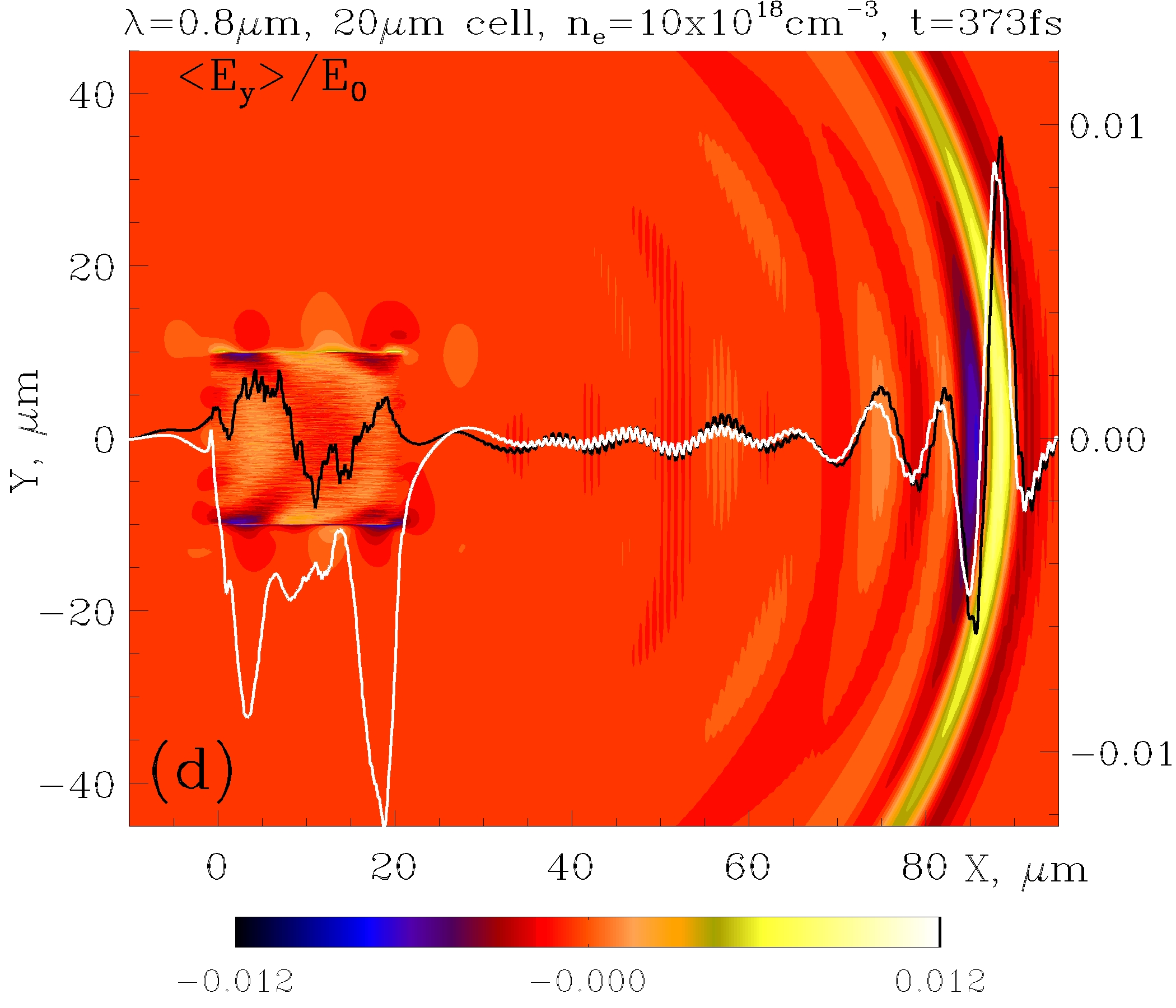}
\includegraphics[width=0.33\textwidth]{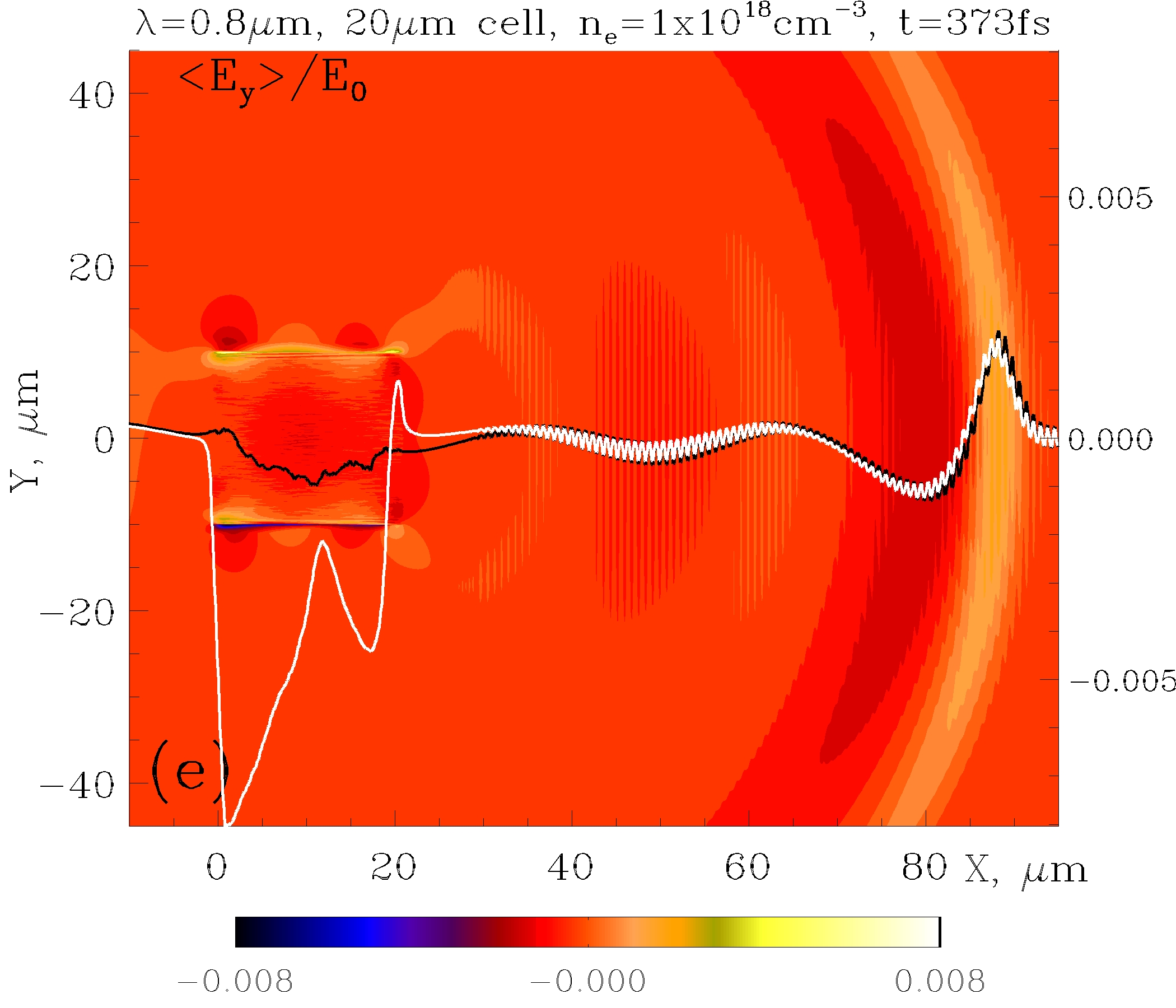}
\includegraphics[width=0.33\textwidth]{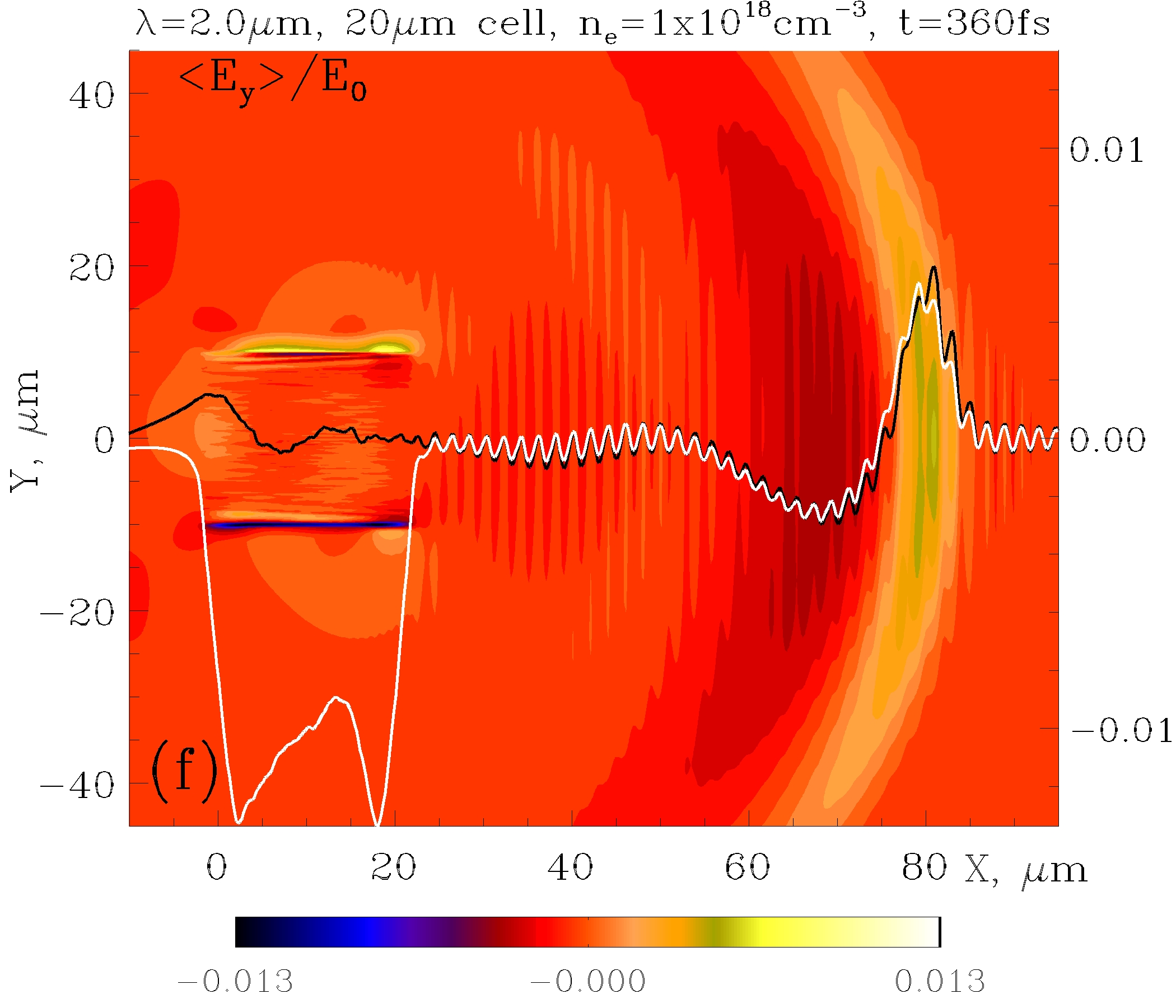}
}
\end{center}
\caption{(color online) Distributions of the time-averaged electric field $E_y(x,y)$ and cuts of these distributions taken on the cell axis $y=0$ (black lines) and the cell boundary, $y=L/2$ (white lines) at times $t\simeq 400~\fs.$ The cell size, atomic concentration, laser wavelength and time instants $t$ are shown on the panels headlines. 
}
\label{fig:fig2}
\end{figure}

Below we focus on the distribution of the electric field inside the plasma and on the total energy emitted in the THz domain.
Analysis of the magnetic fields and of the temporal structure of THz pulses will be presented elsewhere.
Figure \ref{fig:fig2} shows spatial distributions of the electric field $E_y$ taken at times $t\approx 400~\fs$ after the interaction has started.
The fields are averaged over the fundamental period aiming to smear out fast oscillations induced by the two-color pump pulse.
Cuts of $E_y$ alog the $x$-axis for $y=0$ and $y=L/2$ are depicted by black and white lines, respectively.
The plots allow estimating the peak values of the oscillating electric fields inside the plasma and in the near-field zone.
On the cell axis, the magnitude of this quasi-static electric field reaches $E_\mathrm{m}\approx 0.01E_0=8.6~\mathrm{MV/cm}$ for the $50~\mum$ cell at concentration $n_0=10^{19}~\cmcmcm$ and $0.8~\mum$ wavelength.
For the $20~\mum$ cell and $2~\mum$ wavelength the peak electric field inside the plasma is even higher despite of the $10$-times lower electron density.

Finally, we compute the total energy emitted in the THz domain.
To this end, we notice that at $t>400~\fs$ THz pulses are already well localized in space relatively far from the plasma, so that their linear energy density (i.e. energy per unit length in $z$ direction) can be estimated as
\beq
W_{\rm THz}\approx\int\limits_{S}dxdy\frac{{\bar{\vecE}}^2}{4\pi}~
\label{W}
\eeq
where the integral is taken over the 2D area $S$ in the $(x,y)$ plane where the THz wave is localized. 
Here $\bar{\vecE}$ denotes the electric field averaged over fast oscillations with frequencies $\omega$ and $2\omega$.
The averaging procedure rules out the contribution of the pump laser field.
Results are summarized in Table~\ref{table:table1} for the same set of parameters as used for Fig.~\ref{fig:fig2}.

\begin{table}[h]
\caption{\label{table:table1} THz pulse energy per unit length calculated from Eq.~\eqref{W} for parameters of Fig.~\ref{fig:fig2}, and normalized to its value for $n_0=10^{19}~\cmcmcm$, $\lambda=0.8~\mum$ and $L=50~\mum$.}
\begin{center}

\begin{tabular}{llll}
\br
Cell size & $n_0=10^{19}~\cmcmcm$ &  $n_0=10^{18}~\cmcmcm$ &  $n_0=10^{18}~\cmcmcm$\\
~~~ & $\lambda=0.8~\mum$ & $\lambda=0.8~\mum$ & $\lambda=2~\mum$\\
\mr
$50\mu$m & 1.0 & 0.21 & 2.0 \\
$20\mu$m & 0.44 & 0.16 & 0.88 \\
\br
\end{tabular}
\end{center}

\end{table}

\section{Discussion and outlook}
Results reported in the previous section show several clear tendencies.
First, in all cases the THz pulse duration is shorter than that of the pump pulse.
The main spike of the field lasts about $100~\fs$ for $n_0=10^{19}~\cmcmcm$ and $150~\fs$ for $n_0=10^{18}~\cmcmcm$.
These times are approximately equal to the plasma periods $\ds T_p=2\pi/\omega_p$ at the electron concentration equal to $n_0$.
At the same time, the THz pulse duration appears almost independent on the cell size and the pump wavelength.
The short duration of the THz burst contrasts with the fact that plasma oscillations are clearly present in the cell long after the pump laser pulse is gone.
A possible mechanism, which limits the pulse duration, can be connected to an extremely strong radiation damping of the plasma oscillations.
In the long wavelength limit, the radiation reaction force acting on an electron is proportional to the number of coherently radiating electrons.
For THz wavelengths and plasma concentrations considered here, this number can be as large as $N_e\simeq 10^{10}$ leading to an almost prompt damping of that component of the plasma oscillation, which emits radiation.
The following plasma evolution remains time-dependent but does not lead to significant emission anymore.
This behavior can be interpreted as resulting from a weak coupling of the survived plasma oscillation mode to the field of radiation \cite{berge} or, equivalently, as a formation of a specific non-radiating dynamic configuration.
The latter has been extensively studied in nanoplasmonics and optics of metamaterials \cite{nano}.
A quantitative description of this radiation damping effect on the plasma dynamics at long wavelengths will be given elsewhere.

Second, the quasi-static electric field excited inside and in a close vicinity of the cell reach the value $E_\mathrm{m}\approx 0.013E_0=11.6~\mathrm{MV/cm}$ for parameters of Fig.~2(d). 
The peak amplitude practically does not depend on the cell size and grows with the electron concentration and the laser wavelength. 
Assuming a homogeneous plasma oscillation with electron velocity distribution shown on Fig.~\ref{fig:fig1}(c-d), one may estimate the electric field amplitude as $\ds E_\mathrm{m}\simeq E_0 v_0/v_p\sim\sqrt{n_e}\lambda$ where $v_0$ is the characteristic value of the initial velocity $v_y$ and $\ds v_p=eE_0/m_e\omega_p$.
Estimating $v_0$ from the data of Fig.~\ref{fig:fig1} one obtains values of the electric field amplitude close to those shown on Fig.~\ref{fig:fig2}.
Thus, the quasi-static electric field induced inside the cell by the plasma oscillation grows with the atomic concentration and the laser wavelength, the latter agrees with the known data and calculations of two-color THz emission in mid-infrared fields \cite{clerici-prl13,fedorov-pra18,tulsky-pra18}. 
The dependence of the electric field distribution on the cell size is less trivial. At $50~\mum$ the field is mostly concentrated near the cell edges where it oscillates spatially in the $y$-direction with a relatively small period.
These lateral oscillations can be explained as resulting from an non-homogeneous ionization on the shoulders of the pump laser pulse whose width in the transverse direction is determined by the parameter $R=16~\mum$ and is smaller than the cell size.
Instead, for the smaller cell the field ionization is quickly saturated in the whole volume making the system behaving similarly to a plain capacitor with an almost homogeneous time-dependent electric field inside.

Finally, we consider the energy emitted in the THz domain.
Its relative value given in Table 1 shows that the smaller cell emits with a much higher efficiency per electron than that of $50~\mum$ size.
Indeed, taking into account that for our 2D calculation the linear concentration of electrons in the $20~\mum$ cell is $6.25$ times smaller, we obtain, from the numbers shown in Table~\ref{table:table1}, that this cell emits $2.7-4.7$ times more energy per electron, at other parameters fixed.
This enhanced efficiency is the result of higher coherency of the electron motion inside the smaller cell.
While the $50~\mum$ cell supports $2-3$ plasma wavelengths, the smaller cell shows almost a pure dipole oscillation of the electron gas.
As a result, the electrons emit in-phase maximizing both the emission power and the radiation damping.
The higher radiation damping is also responsible for a much weaker tail of the THz pulse in the case of the smaller cell.
For a 3D plasma this quasi-dipole oscillation would make THz emission almost isotropic in the $(x,z)$ plane.
In our 2D model, a considerable emission happens in the backward direction.
Its contribution is not seen on Figs.~\ref{fig:fig2}(e,f) for  $t\simeq 400~\fs$ because the back propagated THz wave at this time is located at $x\approx -80\div -60~\mum$.
Backward THz emission from two-color ionization of air has been recently observed in experiment \cite{bukin-apl19} where the laser pulse was tightly focused to create a small spike instead of a conventional elongated filament.

Effects considered in this paper essentially depend on the plasma size (particularly on that in the longitudinal direction).
A cell restricted by sharp boundaries is a numerically convenient model, but it might appear a setup difficult for experimental realizations.
The problem of cell material, which will unavoidably interact with a strong laser pulse, becomes particularly severe for the smallest considered cell size of $20~\mum$.
Alternative realizations of the scheme with a spatially restricted interaction volume can be achieved by using either tight focusing or thin gas jets.
In the tight focusing scheme, the lateral plasma size will also be limited to the value of several microns, highly reducing the total THz energy output.
Instead, in a gas jet flowing from a thin linear nozzle, the longitudinal and lateral plasma dimensions can be independently controlled by the jet thickness and the focal spot size.

In conclusion, we studied emission of THz waves from a small gas cell where a non-equilibrium plasma is created by means of strong-field two-color ionization of atoms.
Our main funding is the high relative efficiency of the electron energy conversion into THz radiation for smaller plasma emitters. 
Quasi-static electric fields are shown to achieve values $>10~\mathrm{MV/cm}$ and can be further enhanced by applying longer mid-infrared wavelengths and using higher concentrations in the target.
Consequently, a gas jet, gas-filled cell or fiber of size of $10-20~\mum$ can serve as a highly efficient source of quasi-static electric and magnetic fields both in the near-filed and far-field zones.

\section*{Acknowledgments}
Authors acknowledge useful discussions with V. Bukin, A. Gopal, G.G. Paulus and V. Strelkov.
SVP acknowledges financial support of the Russian Science Foundation through grant No.18-12-00476 in the part related to the development of the model and analysis of numerical results. 
The development of numerical algorithms was partially supported by the Russian Science Foundation through the grant No.19-71-20026. 
TVL acknowledges  the Siberian Supercomputer Center ICMMG SB RAS (Novosibirsk) for providing the computing time on NKS-1P cluster. 
Most of the numerical simulations were performed using the computing resources granted by the John von Neumann-Institut f\"ur Computing (Research Center J\"ulich) under the project HRO04.

\end{document}